\documentstyle[multicol,aps,pra,epsfig,floats,here,amstex]{revtex}

\title{Quantum information distribution via entanglement}
\author{Mio Murao$^{1,2}$, Martin B. Plenio$^{2}$ and Vlatko Vedral$^3$}
\address{
$^1$Semiconductor Laboratory, The Institute of Physical and
Chemical Research (RIKEN), Wako-shi 351-01, Japan\\ $^2$Optics
Section, Blackett Laboratory, Imperial College, London SW7 2BZ,
United Kingdom\\ $^3$Centre for Quantum Computing,  Clarendon
Laboratory, University of Oxford, Oxford OX1 3PU, United Kingdom}

\date{\today}

\begin{document}
\draft
\maketitle
\begin{abstract}
We present a generalization of quantum teleportation that
distributes quantum information from a sender's $d$-level particle
to $N_o$ particles held by remote receivers via an initially
shared multiparticle entangled state.  This entangled state
functions as a multiparty quantum information distribution channel
between the sender and the receivers.  The structure of the
distribution channel determines how quantum information is
processed. Our generalized teleportation scheme allows multiple
receivers at arbitrary locations, and can be used for applications
such as optimal quantum information broadcasting, asymmetric
telecloning, and quantum error correction.
\end{abstract}

\pacs{PACS numbers: 03.67.-a, 03.67.Hk}

\section{Introduction}

In the quantum teleportation scheme \cite{Bennett1}, quantum
information of an unknown state of a $d$-level particle (an
``input'' particle) is faithfully transmitted from a sender
(Alice) to a remote receiver (Bob) via an initially shared pair of
maximally entangled particles. The distributed entangled particles
shared by Alice and Bob function as a quantum information channel
for the faithful transmission. Quantum teleportation has been
demonstrated in several successful experiments \cite{experiment}.
It represents the basic building block of future quantum
communication networks between distant parties \cite{channel}.

In addition to the ``one-to-one'' quantum communication of
teleportation, it is natural to consider ``one-to-many'' quantum
communication via quantum channels, i.e. quantum broadcasting from
a sender to several spatially separated receivers.  However, it is
not possible to perform ``one-to-many'' quantum communication
perfectly, because the no-cloning theorem \cite{Zurek} forbids
{\it perfect} duplication of quantum information.  Approximate
methods for quantum cloning are known but these methods require
all parties (the original and all the approximate copies) to be in
one place.

Our strategy for ``one-to-many'' quantum communication is to {\it
distribute} quantum information of a particle from a sender to
many distant receivers.  Such a strategy, dubbed quantum
telecloning, has been suggested in \cite{Murao}.  In the quantum
telecloning scheme, information of an input qubit (a $d=2$-level
particle) is distributed into $M$ particles which are {\it
optimal} clones and $M-1$ which are ancilla particles, all
spatially separated from each other.  This transmission is
achieved by first establishing a particular initial entangled
state between the sender and receivers.  The protocol of quantum
telecloning is then similar to that for original quantum
teleportation \cite{Bennett1}, consisting of a joint measurement
by Alice, 2-bit classical communication from Alice to Bob and a
local operation by Bob.  This ``optimal broadcasting'' of quantum
information relies on the structure of the distributed
entanglement which functions as a one to many quantum
communication channel. Recently, the telecloning protocol has been
generalized to the case of $N \left( \leq M \right)$ identical
input qubits being distributed to $M$ spatially separated parties
by D{\"u}r \cite{Dur}. In this generalization, the same entangled
state of \cite{Murao} is used for the quantum channel, but a more
generalized POVM is performed for the joint measurement.

We also consider variation of this distribution method for
alternative applications. In this paper, we present optimal
quantum information broadcasting for $d$-level particles,
asymmetric telecloning of qubits, and quantum error correction via
entanglement as examples of a generalization of quantum
teleportation to one-to-many quantum communication.  The important
rule of our game is that the receivers are spatially separated
from each other so that we do not allow any global operations
among receivers.

There is an alternative trivial way to distribute quantum
information from a sender to many receivers if we allow the sender
to run quantum networks that involve global operations of many
particles.  In this case, the sender first performs quantum
networks for encoding one particle information into several
particles in her site.  Then she transmits the encoded particles
to each receiver using the original teleportation scheme with two
particle maximally entangled state \cite{Bennett1}.  For the
transmission, $M \log_2 d$ e-bits of entanglement are required to
distribute quantum information of a $d$-level particle into the
spatially separated $M$ receivers. The sender performs the
measurement $M$ times and uses $M d^2/2$ bits of classical
communication from the sender to receivers. On the other hand, in
our direct information distribution scheme, multiparticle
entanglement is used simultaneously for both encoding information
and transmission. We need only a single joint measurement and
require only $d^2/2$ bits of classical communication (announced
publically to all the receivers).  The amount of entanglement
between the sender and the receiver as the whole is $\log_2 d$
e-bits.  Thus our direct information distribution via entanglement
is more efficient in terms of local and global operations,
classical communication and the resource of entanglement.

Distributing information to several different parties can be
useful for protecting against eavesdropping.  Thus the information
distribution can be used for more secure quantum communication.
The ``tele-error-correction'' scheme will provide us with another
interpretation of quantum error correction and an interesting
observation about entanglement required for the quantum channel
for encoding.  Our information distribution scheme can be also
functions as a ``ready made network'' \cite{Popescu} when all the
particles of the quantum channel are at one site, say Alice's
site. The quantum channel is then a ``black box'' having an input
port and several output ports to encode a single particle state
into a multiparticle state.  The ``manufacturer'' performs the
complicated quantum operations to produce the black box. Alice,
the ``user'', only needs to perform the joint measurement for
inputting information, and single particle operations depending on
the measurement outcome, instead of several CNOT operations
required for the global operations of particles.

The rest of the paper is organized as follows. In section
\ref{sec:general} we present our generalization of quantum
teleportation for information distribution. Applications of the
information distribution scheme are then presented in the
following sections. The optimal quantum information broadcasting
($1 \rightarrow M$ telecloning) for $d$-level particles is shown
in section \ref{sec:broadcasting} the asymmetric telecloning for
distributing information with different fidelity for each receiver
are investigated in section \ref{sec:asymmetric}.
Tele-error-correction, quantum error correction via the
information distribution scheme, is presented in section
\ref{sec:tec}. A summary is given in section \ref{sec:summary}.

\section{A generalization of quantum teleportation} \label{sec:general}

In the quantum teleportation scheme of Bennett et al
\cite{Bennett1}, a pair of maximally entangled particles
($d$-level particles) initially shared by a sender (Alice) and a
receiver (Bob) functions as a channel for quantum information with
the help of a classical information channel.  Alice's particle is
used as a ``port'' for information input, and Bob's particle is
used as the ``output'' port in the scheme. We may imagine that
there are two processes taking place during the faithful
transmission of quantum information of an unknown state of an
``input particle'' from Alice to Bob. The first process is the
``information input'' process.  Alice performs a joint measurement
in the maximally entangled basis of the input particle and her
port particle.  Alice obtains one of the $d^2$ possible
measurement results.  This operation ``injects'' quantum
information from the input particle into the quantum channel.  We
call this measurement a ``Bell-type measurement'' not only in the
context of qubits ($d=2$), but in general for $d$-level particles
and the maximally entangled basis is called the Bell basis.
Injected information appears at Bob's output particle as one of
the $d^2$ orthogonal states depending on the result of the
Bell-type measurement. Without information about Alice's
measurement result, the output particle of Bob is in an equal
mixture of $d^2$ orthogonal states, which does not provide any
information about the original states. (If the output state of Bob
gave any information of Alice's input before receiving the
measurement result from Alice, Alice and Bob could communicate
faster than light!)  Thus we need a second process, the ``recovery
unitary operation'' (RUO).  In this process, Alice notifies which
of the $d^2$ possible measurement results she obtained.  Then Bob
performs a unitary operation on the output particle depending on
the measurement result to recover the quantum information of the
input particle.

Now we generalize the quantum teleportation scheme for
distributing quantum information of a $d$-level particle to more
than one receiver via a multiparty quantum channel.  The quantum
channel is a multiparticle entangled state initially shared
between the sender and the receivers (Bob, Charlie, and so on).
The sender and receivers are spatially separated from each other
and no global operation between particles held by different
receivers is allowed.

In our scheme, the input, port and output particles of the
original teleportation scheme are replaced by groups of input
particles, port particles, and output particles.  We represent the
number of the input, port and output particles as $N_i$, $N_p$ and
$N_o$, respectively. We denote the basis of the $d$-dimensional
subspace of the original quantum state (of a $d$-level particle)
as $\left \{ \left \vert \psi_j \right \rangle \right \}$ for the
input particles, $\left \{ \left \vert \pi_j \right \rangle \right
\}$ for the port particles, and $\left \{\left \vert \phi_j \right
\rangle \right\}$ for the output particles, where $j=0,1,...,d-1$.
All these bases are represented by the states of (multi)particles.
For example, information of a $d$-level particle implemented in
the sender's $N_i$ input particles is represented as
\begin{eqnarray}
\left \vert \psi \right \rangle
=
\sum_{j=0}^{d-1}
{a_j \left \vert \psi_j \right \rangle}
\end{eqnarray}
under the constraint $\sum{\left \vert a_j \right \vert^2}=1$.

The quantum channel for information distribution between the
sender and the receivers is a maximally entangled state of the
sender's port particles and the receivers' output particles
\begin{eqnarray}
\left \vert \xi \right \rangle
=
\frac{1}{\sqrt{d}}
\sum_j{\left \vert \pi_j \right \rangle
\otimes
\left \vert \phi_j \right \rangle}.
\label{eqn:qbc}
\end{eqnarray}
The joint state of the input particle and the channel is
\begin{eqnarray}
\left \vert \psi \right \rangle \otimes \left \vert \xi \right
\rangle = \sum_n{\sum_m{ \left \vert \Phi_{nm} \right \rangle
\otimes \frac{1}{\sqrt{d}} \sum_j^{d-1} { \exp \left [ -2 \pi i j
n /d \right ] \cdot \alpha_j \left \vert \phi_{\overline{j+m}}
\right \rangle} }} \label{eqn:channel2}
\end{eqnarray}
where $\overline{j+m} = \left[\left(j+m \right)~{\rm mod}~d
\right]$ and $\left \vert \Phi_{nm} \right \rangle$ is a joint
state of the input particles and the port particle in a maximally
entangled basis (the Bell-type basis)
\begin{eqnarray}
\left \vert \Phi_{nm} \right \rangle
=
\frac{1}{\sqrt{d}}
\sum_k^{d-1}
{ \exp \left [ 2 \pi i k n /d \right ]
\cdot \left \vert \psi_k \right \rangle
\otimes
\left \vert \pi_{\overline {k+m}} \right \rangle
}
\end{eqnarray}
for $0 \le n,m \le d-1$.  Therefore, the RUO for a Bell-type
measurement outcome $\left \vert \Phi_{nm} \right \rangle$ is
given by
\begin{eqnarray}
U_{nm}
=
\sum_j {\exp \left [2 \pi i j n /d \right] \cdot \left \vert
\phi_j \right \rangle \left \langle \phi_{\overline{j+m}} \right
\vert}. \label{eqn:ruo}
\end{eqnarray}

So far we have just relabeled the basis of the input and output
particles from the single particle computational basis $\left
\vert j \right \rangle$ into the multiparticle basis $\left \vert
\psi_j \right \rangle$, $\left \vert \pi_j \right \rangle$ and
$\left \vert \phi_j \right \rangle$.  It is remarkable in our
generalization for quantum information distribution that the RUO
$U_{nm}$ can be replaced by a {\it local} recovery unitary
operation (LRUO) $U_{nm}^{\rm local}$, which is a direct product
of local operations for each particle
\begin{eqnarray}
U_{nm}^{\rm local}={\cal U}_{nm}^1 \otimes \cdots \otimes {\cal
U}_{nm}^{N_o} \label{eqn:lruo}
\end{eqnarray}
where ${\cal U}_{nm}^l$ denotes the local operation for the $l$-th
particle of the receivers.  Although $U_{nm}^{\rm local} \neq
U_{nm}$ in general for the full Hilbert space for the $N_o$
particles, $U_{nm}^{\rm local}$ operates in the same way as the
RUO $U_{nm}$ in Eq.~(\ref{eqn:ruo}) on the {\it subspace} spanned
by the output state basis $\left \{ \left \vert \phi_j \right
\rangle \right \}$
\begin{eqnarray}
U_{nm}^{\rm local} \left \vert \phi_j \right \rangle =U_{nm} \left
\vert \phi_j \right \rangle
\end{eqnarray}
for any $j$. We note that the LRUO Eq.~(\ref{eqn:lruo}) is not always
determined uniquely for the corresponding (global) RUO defined by
Eq.(\ref{eqn:ruo}).  The condition that the RUO is local places
additional constraints on the output state basis.

Since the RUO $U_{nm}$ can be decomposed into the products of
$U_{01}$ and $U_{10}$ from the definition of Eq.~(\ref{eqn:ruo}),
the LRUO $U_{nm}^{\rm local}$ may be decomposed in the similar
manner:
\begin{eqnarray}
U_{nm}^{\rm local}=\underbrace{U_{10}^{local}
 \cdot \ldots \cdot U_{10}^{local}}_{\mbox{$n$ times}} \cdot
\underbrace{U_{01}^{local} \cdot \ldots \cdot
U_{01}^{local}}_{\mbox{$m$ times}}.
\end{eqnarray}
Then the condition for the output state basis is the existence of the
following two LRUOs:
\begin{eqnarray}
U_{01}^{local}
\left \vert \phi_j \right \rangle
={\cal U}_{01}^1 \otimes \cdots \otimes {\cal U}_{01}^{N_o}
\left \vert \phi_j \right \rangle
=
\left \vert \phi_{\overline{j-1}} \right \rangle,
\end{eqnarray}
which changes the state from $\left \vert
\phi_j \right \rangle$ to $\left \vert \phi_{\overline{j-1}}
\right \rangle$,
and
\begin{eqnarray}
U_{10}^{local} \left \vert \phi_j \right \rangle &=&{\cal
U}_{10}^1 \otimes \cdots \otimes {\cal U}_{10}^{N_o} \left \vert
\phi_j \right \rangle \nonumber \\ &=& \exp \left [2 \pi i j n /d
\right] \cdot \left \vert \phi_j \right \rangle,
\end{eqnarray}
which changes the phase depending on the state $\left \vert
\phi_j \right \rangle$.

The protocol for distributing quantum information from a sender
to spatially separated receivers
\begin{eqnarray}
\left \vert \psi \right \rangle_{\rm sender}
=
\sum_{j=0}^{d-1} {a_j \left \vert \psi_j \right \rangle}
\longrightarrow \left \vert \phi \right \rangle_{\rm receivers}
=\sum_{j=0}^{d-1} {\alpha_j \left \vert \phi_j \right \rangle}.
\end{eqnarray}
via the quantum channel defined by Eq.~(\ref{eqn:qbc}) is the following:
\begin{enumerate}
\item{The sender performs a Bell-type measurement on the input
particles and the port particles in the basis $\left \{\left \vert
\Phi_{nm} \right \rangle \right \}$ .  We expect $d^2$ different
measurement outcomes.}
\item{The sender classically (and publicly) broadcasts the measurement
outcome (on which basis of $\left \vert \Phi_{nm} \right \rangle$ she
obtained by the projection) to the receivers.}
\item{Depending on the broadcast result $\left \vert \Phi_{nm} \right
\rangle$, the receivers perform the LRUO $U_{nm}^{\rm local}$. }
\end{enumerate}

Information of an initial state $\left \vert \psi \right \rangle$ (of
d-level system) is faithfully transmitted via the quantum channel to
an encoded state $\left \vert \phi \right \rangle$ which is the state
of the particles distributed among $N_o$ spatially separated
receivers.  This ``teleportation'' of quantum information of a
$d$-level particle is faithful because the channel represented by
Eq.~(\ref{eqn:qbc}) has $\log_2 d$ e-bit entanglement between the
sender and the receivers as the whole.  The appearance of the output
information at each receiver's particle is the result of the
information distribution.  Distribution of information depends on
properties of the output state basis $\left \vert \phi_j \right
\rangle$.

This generalization of quantum teleportation looks simple in this
representation.  However as we will show later, it has more
applications.  Optimal quantum information broadcasting and
asymmetric telecloning are just special cases of the scheme. Also
quantum error correction can be carried out via entanglement with
additional conditions on the output state and a slight extension
of the concept of the Bell-type measurement in the decoding
process.

\section{Optimal broadcasting for multi-level particles}
\label{sec:broadcasting}

Although information in an unknown quantum state cannot be copied
perfectly (no-cloning theorem) \cite{Zurek}, a way has been found
to obtain ``optimal'' copies of the original state by an global
unitary transformation involving several particles \cite{Buzek0}.
The optimality of copies is defined by ensuring the largest
fidelity from the original state.  This quantum optimal cloning of
qubits ($d=2$ particles) has been studied in
\cite{Buzek0,Buzek1,Gisin,Bruss}. While optimal cloning
transformations involve global operations on qubits, we have
recently considered the problem of quantum ``telecloning'' for
qubits ($d=2$) in \cite{Murao}. Telecloning is a combination of
the universal optimal cloning and quantum teleportation performed
simultaneously.  The aim of telecloning is to broadcast
information of an unknown state from a sender to several spatially
separated receivers exploiting an entangled state as a quantum
channel.  The properties of the quantum channel for the qubit
telecloning has been investigated in \cite{Murao}.

For the more general case, the problem of optimal cloning of $N$
identical unknown input states to $M$ output copies of $d$-level
particles, which is called ``$N \rightarrow M$ optimal quantum
cloning'' is formulated in \cite{Werner}.  In that paper, Werner
has shown that the optimal cloning map ${\hat T}$ to obtain $M$
optimal clones from $N$ identical (unknown) input states is the
projection of the direct product of the $N$ input states and $M-N$
{\it identity} states onto the symmetric subspace of $M$
particles:
\begin{eqnarray} \label{eqn:cloningmap}
{\hat T} \left ( \rho \right )
=
\frac{d \left[ N \right ]}
{d \left[ M \right]}
s_M \left( \rho \otimes {\boldmath{1}}^{\otimes \left(
M-N \right)} \right) s_M
\end{eqnarray}
where $s_M$ is the projection operator for the symmetrized state
of $M$ $d$-level particles, $\rho$ is the density operator for the
input state given by the direct product of an input state $\left
\vert \psi \right \rangle \left  \langle \psi \right
\vert$
\begin{eqnarray}
\rho=\underbrace{\left \vert \psi \right \rangle \left  \langle \psi \right
\vert \otimes \cdots \otimes \left \vert \psi \right \rangle \left
\langle \psi \right \vert}_{\mbox
{$n$ times}},
\end{eqnarray}
and $d \left[N \right]$ is the number of the symmetrized state for
$N$ $d$-level particles given by $d \left[ N \right]= _{d+N+1}{\rm
C}_{N}$.  Werner has proved the optimality of the clones obtained
by the cloning map Eq.~(\ref{eqn:cloningmap}) mathematically.  The
cloning map is the completely positive, trace preserving map.
However, physical implementation of the optimal cloning map is not
obvious, since the cloning map (\ref{eqn:cloningmap}) is not a
unitary transformation.  To find the corresponding unitary
transformation for the optimal cloning of $d$-level particles, we
need a pure state representation including ancilla particles.

In this section, we present optimal quantum information
broadcasting for multi-level particles ($1 \rightarrow M$
telecloning of a $d$-level particle) , as an application of the
quantum information distribution scheme described in the previous
section. In our scheme, we obtain the pure state representing $M$
optimal clones of an unknown state and $M-1$ ancillas.
Consequently, we find the unitary transformation which implements
the Werner's optimal cloning map for $d$-level particles
\cite{Werner}.

For optimal quantum information broadcasting, the input state
basis and the port state basis are taken in a single particle
computational basis $\left \{ \left \vert j \right \rangle \right
\}$.  The output state basis is represented by the states $\left
\{ \left \vert \phi_j \right \rangle \right \}$ ($j=0,1,...,d-1$)
consisting of $N_o=2M-1$ $d$-level particles where $M-1$ particles
are ancillas and $M$ particles are for presenting optimal cloning
states at the end of the protocol.

The output state basis consisting of $2M-1$ particles is
represented in terms of the normalized and the symmetrized state
$\left \vert \xi_k^M \right \rangle$ of $M$ $d$-level particles
\begin{eqnarray} \label{eqn:optclonbasis}
\left \vert \phi_j \right \rangle
=\frac{\sqrt{d}}{\sqrt{d \left[ M \right]}}
\sum_{k=0}^{n_M-1}
\underset{P}{} \left \langle j \left \vert \right.
\xi_k^M \right \rangle_{PA} \otimes
\left \vert \xi_k^M \right \rangle_{C}
\end{eqnarray}
where $P$ denotes the port particles, $A$ denotes the $M-1$
ancilla particles and $C$ denotes the $M$ particles for optimal
clones. The structure of the symmetrized state is the key feature
for our optimal quantum information broadcasting as we will show
later. In the computational basis, the symmetrized state $\left
\vert \xi_k^M \right \rangle$ can be represented by
\begin{eqnarray}
\left \vert \xi_k^M \right \rangle
=\frac{1}{\sqrt{{\cal N} \left(\xi_k^M \right)}}
\left \vert
{\cal P} \left(
a_0,a_1,\cdot \cdot \cdot ,a_{M-1}
\right)
\right \rangle,
\end{eqnarray}
where ${\cal P}$ denotes an operator which creates the sum of all
possible states represented by permutation of the elements $\left \{
a_0, \cdots, a_{M-1} \right \}$ for $a_n \in \left \{0,1,\cdot \cdot
\cdot d-1 \right \}$ and $a_{n+1} \ge a_{n}$ under the constraint
${\cal N} \left(\xi_k^M \right)$ imposing the normalization of $\left
\vert \xi_k^M \right \rangle$. The index $k$ for the symmetrized state
is defined by the following: First we assign to each string $\left \{
a_0, \cdots, a_{M-1} \right \}$ a number
\begin{eqnarray}
h \left(a_0,a_1,\cdot \cdot \cdot ,a_{M-1} \right)
=\sum_{n=0}^{M-1} a_n d^{M-1-n}.
\end{eqnarray}
Then we sort those numbers in increasing order.  The index $k$ ($0
\le k \le d \left[ M \right]$) is then associated with the string
$\left \{ a_0, \cdots, a_{M-1} \right \}$
\begin{eqnarray}
k=f_M \left (  a_0, \cdots , a_{M-1}  \right )
\end{eqnarray}
giving rise to the $(k+1)$th smallest number $h \left(a_0,a_1,\cdot
\cdot \cdot ,a_{M-1} \right)$.

The LRUOs for the output state basis $\left \{ \left \vert \phi_j
\right \rangle \right \}$ are given by
\begin{eqnarray}
U_{nm}^{\rm local}= \underbrace{{\cal U}_{nm}^A \otimes \cdots
\otimes {\cal U}_{nm}^A}_{\mbox{$M-1$ ancillas}} \otimes
\underbrace{{\cal U}_{nm}^C \otimes \cdots \otimes {\cal
U}_{nm}^C}_{\mbox{$M$ clones}} \label{eqn:tclro}
\end{eqnarray}
where
\begin{eqnarray}
{\cal U}_{nm}^{A}=
\sum_{j=0}^{d-1} {\exp \left [-2 \pi i j n /d \right] \cdot
\left \vert j \right \rangle \otimes
\left \langle \overline{j+m} \right \vert},
\label{eqn:tclruoa}
\end{eqnarray}
and
\begin{eqnarray}
{\cal U}_{nm}^{C}=
\sum_{j=0}^{d-1} {\exp \left [2 \pi i j n /d \right] \cdot
\left \vert j \right \rangle \otimes
\left \langle \overline{j+m} \right \vert},
\label{eqn:tclruoc}
\end{eqnarray}
which has the complex conjugate phase of Eq.(\ref{eqn:tclruoc}).

The quantum channel is the maximally entangled state between the port
particle and the output state particles
\begin{eqnarray}
\left \vert \xi \right \rangle =
\frac{1}{\sqrt{d}}
\sum_{j=0}^{d-1} \left \vert j \right \rangle_P \otimes
\left \vert \phi_j \right \rangle.
\label{eqn:tcstate2}
\end{eqnarray}
It can also be represented in terms of the symmetrized states as
\begin{eqnarray}
\left \vert \xi \right \rangle =\frac{1}{\sqrt{d \left[ M
\right]}} \sum_{k=0}^{d \left[ M \right]-1} \left \vert \xi_k^M
\right \rangle_{PA} \otimes \left \vert \xi_k^M \right
\rangle_{C}. \label{eqn:tcstate}
\end{eqnarray}
The two groups of particles in the information distribution
channel, the $PA$ group and the $C$ group, are symmetric to each
other.  This symmetry property leads to an LRUO $U_{nm}^{\rm
local}$, which is the product of the local operations given by
Eqs.~(\ref{eqn:tclro})--(\ref{eqn:tclruoc}).  The quantum channel
is a maximally entangled state of $\left( d \left[M
\right]\right)$-level particles between the $PA$ group and the $C$
group.

As we have shown in the previous paper \cite{Murao}, for the case
of $d=2$ (i.e. for qubits), only the $M$ receivers' clone qubits
in the quantum broadcasting channel are ``directly'' entangled to
the port qubit according to the Peres-Horodecki criterion
\cite{Peres-Horodecki}.  If the partial transpose of the density
operator is {\it not} positive, the two particles are entangled
and otherwise, they are disentangled.  The ``structure'' based on
the two particle entanglement of the quantum channel is essential
for the optimal quantum information broadcasting of $d$-level
particles like in the qubit case. However, because the
Peres-Horodecki criterion is only valid for the case of limited
dimensional particles (two qubits, or entanglement of a qubit and
a ``qutrit''), the necessity of the two particle entanglement for
the distribution of quantum information is still a conjecture.

Exploiting the communication channel given by
Eq.~(\ref{eqn:tcstate2}) and following the protocol for the
quantum information distribution described in the previous
section, an unknown input state of the sender
\begin{eqnarray}
\left \vert \psi \right \rangle =
\sum_{j=0}^{d-1} \alpha_j \left \vert j \right
\rangle
\end{eqnarray}
is remotely ``encoded'' into the output state
\begin{eqnarray}
\left \vert \phi \right \rangle &=&
\sum_{i=0}^{d-1} \alpha_j \left \vert \phi_j
\right \rangle.
\label{eqn:optclonestate}
\end{eqnarray}
held by the $2M-1$ specially separated receivers via the quantum
channel.  This output state represents the $M-1$ ancillas and the
$M$ optimal clones.  As we will show in the following, the reduced
density matrix for $M$ optimal clones coincides with a special
case ($N=1$) of the $N \rightarrow M$ universal optimal cloning
state for $d$-level particles, which was proved by Werner
\cite{Werner}.

The key property of the symmetrized state for our proof is that
the symmetric state of $M$ particle can be decomposed into single
particle states and symmetric states of the other $M-1$ particles:
\begin{eqnarray}
\left \vert \xi_k^M \right \rangle
&=&
\frac{1}{\sqrt{{\cal N}\left (\xi_k^M \right)}}
\sum_{a_j \in \left \{0,\cdots,d-1 \right \}}
\left \vert a_j \right \rangle
\left \vert {\cal P}_{M-1} \left
( a_0,\cdots,a_{j-1},a_{j+1},\cdots,a_{M-1} \right ) \right \rangle
\nonumber \\
&=&
\frac{1}{\sqrt{{\cal N}\left (\xi_{k}^{M} \right)}}
\sum_{a_j \in \left \{0,\cdots,d-1 \right \}}
{\sqrt{{\cal N} \left (\xi_{k^\prime}^{M-1} \right)}}
\left \vert a_j \right \rangle
\left \vert \xi_{k^\prime}^{M-1} \right \rangle
\label{eqn:symdecomp}
\end{eqnarray}
where $k^\prime=f_M \left
(a_0,\cdots,a_{j-1},a_{j+1},\cdots,a_{M-1} \right )$.  The sum in
Eq.~(\ref{eqn:symdecomp}) is a special sum, it is taken only for
different values of $a_j \in \left \{0,\cdots,d-1 \right \}$ (if
$a_j=a_{j^\prime}$, only the smaller index $j<j^\prime$ is kept in
the sum).  To make the relationship between the index $k$ and
$k^\prime$ clearer, we define another function $g$ that gives the
index $k$ of the symmetrized state of $M$ particles when a value
of the particle $a_j$ is inserted in the $\left ( j-1 \right )$th
position of a symmetrized state of $M-1$ particle having the index
$k^\prime$:
\begin{eqnarray}
k=g \left(a_j,k^\prime \right).
\end{eqnarray}
Then the output state basis Eq.~(\ref{eqn:optclonbasis}) is
represented by
\begin{eqnarray}
\left \vert \phi_i \right \rangle
=\frac{\sqrt{d}}{\sqrt{d \left[ M \right ]}}
\sum_{{k^\prime}=0}^{d \left[ M-1 \right ]-1}
{\cal R}_i^{k^\prime}
\left \vert \xi_{k^\prime}^{M-1} \right \rangle_A
\otimes
\left \vert \xi_{g \left(i,k^\prime \right)}^{M} \right \rangle_C
\label{eqn:decompxi}
\end{eqnarray}
where
\begin{eqnarray}
{\cal R}_{i}^{k^\prime}=\frac{\sqrt{{\cal N} \left(\xi_{k^\prime}^{M-1} \right )}}
{\sqrt{{\cal N} \left (\xi_{g \left(i,{k^\prime} \right)}^M
\right)}}.
\end{eqnarray}
A detailed derivation of the Eq.~(\ref{eqn:decompxi}) is found in
the appendix.

The reduced density matrix of the clones is obtained by tracing
over the ancilla variables
\begin{eqnarray} \label{eqn:rhoc}
\rho_C&=&{\rm tr}_A \left \vert \phi \right \rangle
\left \langle \phi \right \vert \nonumber \\
&=&
\sum_{l=0}^{n_{M-1}-1} {}_A \left \langle \xi_l^{M-1}
\right \vert \phi \left \rangle
\left \langle \phi \right \vert
\xi_l^{M-1} \right \rangle_A \nonumber \\
&=&\frac{d}{n_M}
\sum_{i=0}^{d-1}
\sum_{i^\prime=0}^{d-1}
\sum_{{k^\prime}=0}^{d \left[ M \right]-1}
\alpha_i \alpha_{i^\prime}^*
{\cal R}_i^{k^\prime} {\cal R}_{i^\prime}^{k^\prime}
\left \vert \xi_{g \left(i,{k^\prime} \right)}^{M} \right \rangle_C
\left \langle \xi_{g\left(i^{\prime},{k^\prime} \right)}^{M} \right \vert.
\end{eqnarray}
The projection operator to the symmetric subspace of $M$ particles
in Werner's cloning map given by Eq.~(\ref{eqn:cloningmap}) in our
notation is
\begin{eqnarray}
s_M=\sum_{k=0}^{d \left[ M \right]} \left \vert \xi_k^M \right \rangle
\left \langle \xi_k^M \right \vert.
\end{eqnarray}
Then the density matrix for $1 \rightarrow M$
$d$-level optimal clones obtained by Werner \cite{Werner} is
represented as
\begin{eqnarray}
\rho_C&=&{\rm {\hat T}} \left (
\left \vert \psi \right \rangle
\left \langle \psi \right \vert
\right ) \nonumber \\
&=&
\frac{n_1}{d \left[ M \right]}
s_M
\left \vert \psi \right \rangle
\left \langle \psi \right \vert
\otimes
{\bf 1}^{\otimes M-1}
s_M \nonumber \\
&=&\frac{d}{d \left[ M \right]}
\sum_{i=0}^{d-1}
\sum_{i^\prime=0}^{d-1}
\sum_{{k^\prime}=0}^{n_{M-1}-1}
\alpha_i \alpha_{i^\prime}^*
{\cal R}_i^{k^\prime} {\cal R}_{i^\prime}^{k^\prime}
\left \vert \xi_{g\left(i,{k^\prime} \right)}^{M} \right \rangle_C
\left \langle \xi_{g\left(i^{\prime},{k^\prime} \right)}^{M} \right
\vert.
\end{eqnarray}
This density matrix coincides with our reduced density matrix for
the clones Eq.~(\ref{eqn:rhoc}).  Thus the output state $\left
\vert \phi \right \rangle$ given by Eq.~(\ref{eqn:optclonestate})
represents the optimal cloning state consisting of $M-1$ ancillas
and $M$ optimal clones.

\section{Asymmetric telecloning} \label{sec:asymmetric}

Quantum telecloning described in the previous section distributes
information from an input state evenly to distant receivers.
However, it may be desirable to distribute information unevenly to
the receivers.  For example, if we trust Alice more than Bob, we
may decide to distribute more information to Alice. Asymmetric
quantum telecloning distributes information from an unknown input
particle into several different parties with {\it different
fidelity} for each party. The corresponding local operation for
this information distribution is the asymmetric cloning proposed
by \cite{Cerf,Buzek-asymmetric}. In this section, we show an
example of $1 \rightarrow 2$ asymmetric telecloning for qubits
($d=2$).

For asymmetric telecloning, the input state basis is taken as the
one qubit computational basis $\left \{ \left \vert j \right
\rangle \right \}$.  The output basis consist of three qubits, one
ancilla qubit $A$ held by Anne (note: she is not Alice) and two
clone qubits $B$ and $C$ held by the receivers, Bob and Claire, as
\begin{eqnarray}
\left \vert \phi_0 \right \rangle
&=&\frac{1}{\sqrt{\cal N}} \left(
\left \vert 000 \right \rangle +
p \left \vert 101 \right \rangle +
q \left \vert 110 \right \rangle \right) \\
\left \vert \phi_1 \right \rangle
&=& \frac{1}{\sqrt{\cal N}} \left(
\left \vert 111\right \rangle +
p \left \vert 010 \right \rangle +
q \left \vert 001 \right \rangle \right),
\label{eqn:astcbasis}
\end{eqnarray}
where $q=1-p$, ${\cal N}$ is a normalization factor given by ${\cal
N}=1+p^2+q^2$ and the order of the qubits is $\left \{A,B,C \right
\}$.  The LRUOs are given by $U_{01}^{local}=\sigma_z \otimes \sigma_z
\otimes \sigma_z$ and $U_{10}^{local}=\sigma_x \otimes \sigma_x
\otimes \sigma_x$.  The information distribution channel for
asymmetric telecloning, which is a maximally entangled state of the
port qubit and the output basis $\phi_0$ and $\phi_1$, is given by
\begin{eqnarray}
\left \vert \xi \right \rangle
&=&
\frac{1}{\sqrt{2}}\left (\left \vert 0 \right
\rangle \left \vert \phi_0 \right \rangle+\left \vert 1 \right \rangle
\left \vert \phi_1 \right \rangle \right ) \nonumber \\
&=&
\frac{1}{\sqrt{2 \cal N}}
\left \{
\left \vert 0000 \right \rangle +
\left \vert 1111 \right \rangle  \right. \nonumber \\
&+& \left.
p
\left (
\left \vert 0101 \right \rangle +
\left \vert 1010 \right \rangle
\right ) +
q
\left (
\left \vert 0110 \right \rangle +
\left \vert 1001 \right \rangle
\right )
\right \}.
\end{eqnarray}
The channel can be illustrated as follows Fig.\ref{fig:astcstate}
in the case $p > q$:
\begin{figure}[H]
  \begin{center}
    \epsfig{file=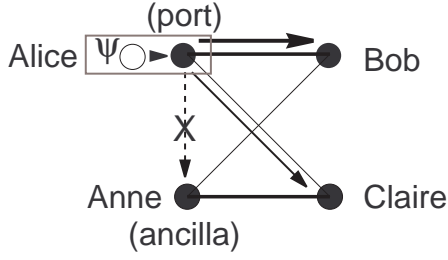,width=2.5truein}
  \end{center}
\caption{An asymmetric telecloning state. The width of lines between
two particles represents the ``strength'' of entanglement between the
two particles.  The difference of strength of entanglement causes
asymmetric telecloning.}
\label{fig:astcstate}
\end{figure}

The information distribution channel for symmetric telecloning is,
of course, given by the choice of parameters $p=q=1/2$.  If we
choose $p=0$ or $q=0$, the asymmetric telecloning state consist of
two maximally entangled pairs (EPR pairs).  In this case, the
receiver who is sharing the EPR pair with Alice obtains faithful
information of the input state and the other who is sharing the
EPR pair with Anne, obtains no information at all.  We now
investigate how the parameters control the asymmetric distribution
of quantum information via entanglement.

Our generalized teleportation protocol with the choice of the
distribution channel (\ref{eqn:astcbasis}) maps the unknown input
state $\left \vert \psi \right \rangle=\alpha_0 \left \vert 0 \right
\rangle + \alpha_1 \left \vert 1 \right \rangle$ to the three qubit
state
\begin{equation}
\left \vert \phi \right \rangle_{ABC} =
\alpha_0 \left \vert \phi_0 \right \rangle +
\alpha_1 \left \vert \phi_1 \right \rangle.
\end{equation}
The asymmetric clones are represented by the reduced density
matrices,
\begin{eqnarray}
\rho_{B}&=&{\rm tr}_{AC} \left \vert \phi \right \rangle
\left \langle \phi \right \vert \nonumber \\
&=&
\frac{1+p^2}{{\cal N}} \vert \psi \rangle \langle \psi \vert
+\frac{q^2}{{\cal N}} \vert \psi^\perp \rangle \langle \psi^\perp \vert
\end{eqnarray}
for Bob's clone and
\begin{eqnarray}
\rho_{C}&=&{\rm tr}_{AB} \left \vert \phi \right \rangle
\left \langle \phi \right \vert \nonumber \\
&=&
\frac{1+q^2}{{\cal N}} \vert \psi \rangle \langle \psi \vert
+\frac{p^2}{{\cal N}} \vert \psi^\perp \rangle \langle \psi^\perp \vert
\end{eqnarray}
for Claire's clone, where $\vert \psi^\perp \rangle$ represents a
state orthogonal to the input state $\vert \psi \rangle$.

To investigate the structure of the quantum channel for asymmetric
telecloning based on two particle entanglement, we calculate the
Peres-Horodecki criterion \cite{Peres-Horodecki}.  For asymmetric
telecloning, the Peres-Horodecki criterion for the reduced density
matrix of the port qubit and Bob's qubit (for the asymmetric
clone) $\rho_{PB}$ is
\begin{eqnarray}
c_B \left(p \right)=\frac{1- 4p+p^2}{4 \left( 1-p+p^2\right)}
\end{eqnarray}
and that for  the reduced density matrix for the port qubit and
Claire's qubit $\rho_{PC}$ is
\begin{eqnarray}
c_C  \left(p \right)=\frac{-2+2p+p^2}{4 \left( 1-p+p^2\right)}.
\end{eqnarray}
There is an interesting case, $c_C=0$, which is given for the
parameter $p=\sqrt{3}-1$.  In this case, the port qubit and the
clone qubit of Claire are not directly entangled with each other
and the fidelities of the clones, which is the matrix element of
the reduced density matrix in terms of the input state $\left
\langle \psi \right \vert \rho_B \left \vert \psi \right \rangle$,
are
\begin{eqnarray}
f_B&=&\frac{2}{3}+\frac{\sqrt{3}}{6}
\end{eqnarray}
for Bob's asymmetric clone and
\begin{eqnarray}
f_C&=&\frac{2}{3}
\end{eqnarray}
for Claire's asymmetric clone.  The state of fidelity $2/3$ is
obtained in the classical limit \cite{Popescu-Masser}.  That is,
only the ``classical'' information of the input state $\psi$ is
transmitted via this disentangled channel.

Here we note that some ``classical'' information of the input
state has also been transmitted to the ancilla qubits of Anne.
Since the reduced density matrix of the ancilla qubit is given by
\begin{eqnarray}
\rho_{A}&=&{\rm tr}_{BC} \left \vert \phi \right \rangle
\left \langle \phi \right \vert \nonumber \\
&=&
\frac{1}{{\cal N}} \vert \psi \rangle \langle \psi \vert
+\frac{p^2+q^2}{{\cal N}} \vert \psi^\perp \rangle \langle \psi^\perp
\vert,
\end{eqnarray}
the ancilla qubit (after the telecloning protocol) can be
considered to be a ``clone'' of very low quality, the fidelity
$1/{\cal N} \leq 2/3$, where the equality is taken at $p=q=1/2$
(symmetric telecloning).  The ancilla qubits only contains
``classical'' information of the input state.  For the asymmetric
case, the ratio of the fidelity of the clones for Bob, Claire and
Anne is $\left [1+p^2 \right]:\left[1+\left(1-p \right)^2
\right]:1$, and Anne always keeps a ``junk'' clone which only
contains some classical information of the input state
irrespective of the parameter $p$.

As pointed out by D{\"u}r \cite{Dur}, the reduced density matrix of
the symmetric telecloning state of the port and clone qubits $\rho_{P
B}$ is a Werner state $\rho_W$ \cite{Werner}.  A Werner state is a
state which is diagonal in the maximally entangled state basis $\left
\{\Phi^+=\Phi_{00},\Phi^-=\Phi_{01},\Psi^+=\Phi_{10},\Psi^-=\Phi_{11}
\right \}$.  The largest diagonal element of $\rho_{PB}$ (fidelity) is
\begin{eqnarray}
\left \langle \Phi^+ \right \vert \rho_W \left \vert
\Phi^+ \right \rangle
=\frac{3 \left( M+1 \right)}{6 M}.
\end{eqnarray}
Thus if we only ``see'' the port qubit and the one of the receivers'
qubit, $1 \rightarrow M$ (symmetric) quantum telecloning is equivalent
to the standard teleportation using an imperfect quantum channel made
of the Werner state $\rho_W$.  For the case of $M=2$, the fidelity of
the Werner state is $3/4$.

For asymmetric telecloning, the reduced density matrix of the quantum
channel is also represented by the Werner state as
\begin{eqnarray}
\rho_{PB}&=&\frac{1}{2 {\cal N}} \left \{
\left(1+p \right)^2 \right.
\left \vert \Phi^+ \right \rangle \left \langle \Phi^+ \right \vert
\nonumber \\
&+& \left.
q^2
\left (
\left \vert \Phi^- \right \rangle \left \langle \Phi^- \right \vert+
\left \vert \Psi^+ \right \rangle \left \langle \Psi^+ \right \vert+
\left \vert \Psi^- \right \rangle \left \langle \Psi^- \right \vert
\right)
\right \}.
\end{eqnarray}
This representation of the quantum channel shows the relation
between the asymmetric telecloning and Cerf's Pauli Cloning
machines \cite{Cerf}. Cerf has suggested that a Pauli Cloning
Machine, which exploits a depolarising channel represented by the
Werner state, performs as a universal asymmetric cloning machine.
The Pauli Cloning Machine is universal (i.e. independent of input
states) only in the case of depolarising channels.  Therefore our
asymmetric telecloning is the only scheme to distribute
information with a given ratio of fidelity irrespective of the
input state.

\section{Tele-error-correction} \label{sec:tec}

Since decoherence is the main obstacle to quantum information
processing, the discovery of quantum error correction schemes
\cite{EC} is very important for the practical realization of
quantum computation and quantum communication.  In this section,
we show how quantum error correction can be performed via
distributed entanglement as another example of our information
distribution scheme.

The standard quantum error correction schemes \cite{EC} consist of
the following four processes.
\begin{enumerate}
\item{The first process involves encoding information. Information of
a qubit is encoded into a state of $N_e$ qubits ($N_e=3$ for the case
that only one kind of error happening to one of the qubits, $N_e=5$ or
$N_e=7$ for the case that one of the three kinds of errors happening
to one of the qubits) by an global unitary transformation of $M$
qubits.}
\item{After encoding, you may have an error in one of the encoded
qubits. The second process is for decoding information of the state
after an error occurs.}
\item{The decoding process is performed by a reverse global unitary
transformation of encoding.  After the decoding process, one of
the qubits is an ``output qubit'' and the others are ancilla
qubits which indicates whether an error occurred. The relationship
of the states of the ancilla qubits and in which qubit the error
occurred is given in the syndrome table.}
\item{The fourth process is to correct errors.
We measure the ancilla qubits and correct an error indicated by
the measurement result and the syndrome table. Alternatively, some
global transformation among the decoded qubits may be performed
for error correction instead of measuring ancilla qubits.}
\end{enumerate}

The first process, encoding qubit information into a state of many
qubits for error correction, is carried out via our information
distribution scheme with the appropriate choice of output state
basis consisting of $N_e$ qubits.  We limit ourselves to the case
of correction of a single error.  Three kinds of errors may happen
to a qubit in the encoded state.  These are equivalent to the
single qubit operation $\sigma_z$ (type $1$), $\sigma_x$ (type
$2$) or $\sigma_z \cdot \sigma_x$ (type $3$).  An error of the
type $l$ ($l=1,2,3$) happening to the $\eta$th particle
($\eta=1,\cdots N_e$), $\epsilon_l^\eta$, maps from an output
state basis $\left \vert \phi_j \right \rangle$ to a state $\left
\vert \varphi_j^\zeta \right \rangle$. The index $\zeta$ is
determined by $\zeta=\left( l-1 \right ) N_e + \eta$.  We define
that $\zeta=0$ represents no error. In some other cases, only one
type of errors is expected.  In this case, we only need to
consider $l=1$.  If we denote the total possible types of errors
as $L$, $\zeta$ takes $\left(L N_e + 1 \right)$ different values,
$\zeta=0,\cdots, L N_e$.  The state changes through the encoding
process and error as follows:
\begin{eqnarray}
\left \vert \psi \right \rangle= \sum_{j=0}^1 \alpha_j \left \vert
j \right \rangle &\longrightarrow& \left \vert \phi \right
\rangle= \sum_{j=0}^1 \alpha_j \left \vert \phi_j \right \rangle
\nonumber \\ &\longrightarrow& \left \vert \varphi \right \rangle=
\sum_{j=0}^1 \alpha_j \left \vert \varphi_j^{\zeta} \right
\rangle.
\end{eqnarray}

For successful information distribution scheme via teleportation,
the output state basis is required to have the LRUO $U_{nm}^{\rm
local}$. In addition, it has to satisfy the following condition
\begin{eqnarray}
\left \langle \phi_{j^\prime} \right \vert {\epsilon_{l^\prime}^{\eta^\prime}}^\dagger
\epsilon_{l}^{\eta} \left \vert \phi_j \right \rangle
=
\left \langle \varphi_{j^\prime}^{\zeta^\prime} \right \vert
\left. \varphi_{j}^{\zeta} \right \rangle
=
\delta_{j,{j^\prime}} \delta_{\zeta,{\zeta^\prime}}
\end{eqnarray}
for error correction.  This condition states that different errors
map a state into different states so that it is possible to
distinguish different errors.  The state after an error indicated
by $\zeta$ is not in the subspace of the original output state
basis $\left \{\left \vert \phi_j \right \rangle \right \}$ but in
the subspace of the $\left \{ \left \vert \varphi_j^\zeta \right
\rangle \right \}$. We now treat the Hilbert space of dimensions
$L N_e +1$, which is the sum of all subspaces for given $\zeta$,
instead of $2$ dimensions for a qubit and the subspace of the
output state.

The decoding and the error correction steps can be described by
the information distribution scheme analogous to telecloning
instead of performing a global unitary operation. We use a pair of
maximally entangled qubits for the quantum channel ($\left \vert
\xi_d \right \rangle=\left( \left \vert 00 \right \rangle+ \left
\vert 11 \right \rangle \right)/\sqrt{2}$), and an ``extended''
Bell-type measurement for the enlarged space occupied by the state
after an error occurs.  The joint state of the error state and the
channel $\left \vert \xi_d \right \rangle$ is
\begin{eqnarray}
\left \vert \varphi \right \rangle \otimes \left \vert \xi_d
\right \rangle = \sum_n{\sum_m{ \left \vert \Phi_{nm}^\zeta \right
\rangle \otimes \frac{1}{\sqrt{2}} \sum_j^{1} { \exp \left [ - \pi
i j n \right ] \cdot \alpha_j \left \vert {\overline{j+m}} \right
\rangle} }},
\end{eqnarray}
where $\left \vert \Phi_{nm}^\zeta \right \rangle$ denotes the
measurement outcomes of the extended Bell type measurement performed
by the sender
\begin{eqnarray}
\left \vert \Phi_{nm}^{\zeta} \right \rangle
=
\frac{1}{\sqrt{2}} \sum_{k=0}^{1} { \exp \left [ \pi i k n \right
] \cdot \left \vert \varphi_k^\zeta \right \rangle \otimes \left
\vert {\overline {k+m}} \right \rangle }
\end{eqnarray}
for $n,m=0,1$ and $\zeta=0,L N_e$.  There are $4 \left(L N_e+1
\right)$ different outcomes possible measured by the extended Bell
measurement.  However we only need information of $n$ and $m$ for
finding the appropriate RUO.  So the sender only needs to broadcast
2-bits of classical information to the receiver.  The RUO for the
output qubit
\begin{eqnarray}
U_{nm} = \sum_{j=0}^1 {\exp \left [\pi i j n \right] \cdot \left \vert
j \right \rangle \left \langle {\overline{j+m}} \right \vert}.
\end{eqnarray}
will give the error corrected original state $\left \vert \psi
\right \rangle$ (in a remote place from the error state).

\subsection{3-qubit code}

\begin{figure}[H]
  \begin{center} \epsfig{file=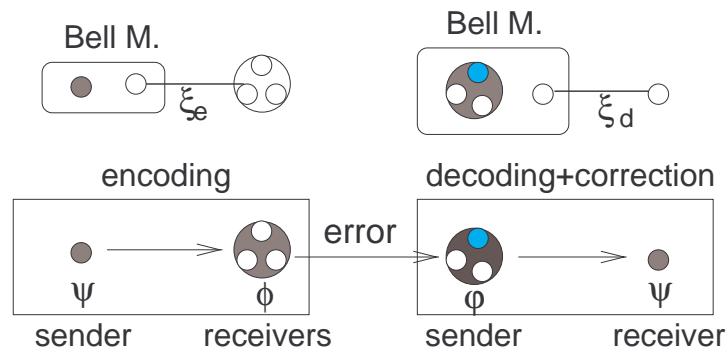,width=10cm} \end{center}
\caption{Amplitude (or phase) error correction via generalized
teleportation is illustrated.  $\xi_e$ denotes the quantum channel
for encoding and $\xi_d$ denotes the quantum channel for decoding
and error correction.  The first raw represents the protocol and
the second raw represents how quantum information is encoded in
each process.} \label{fig:errorcorrection}
\end{figure}

To illustrate our tele-error-correction scheme, we present a
simple example, a three qubit error correction code
(Fig.~\ref{fig:errorcorrection}).  This code is able to correct an
error, which is known to be one of $\left
\{\sigma_z,~\sigma_x,~\sigma_z \cdot \sigma_x \right \}$ that
happens to one of the qubits in the encoded state. In the
following, we investigate the case of an amplitude error (type
$2$). We start from the encoding process. The output state basis
for encoding is the three qubit state:
\begin{eqnarray}
\left \vert \phi_0 \right \rangle &=&\left \vert {0}
{0} {0} \right \rangle \\ \left \vert \phi_1 \right
\rangle &=&\left \vert {1} {1} {1} \right
\rangle.
\end{eqnarray}
The LRUO is given by $U_{01}^{local}=\sigma_x
\otimes \sigma_x \otimes \sigma_x$ and $U_{10}^{local}=\sigma_z
\otimes \sigma_z \otimes \sigma_z$ in the computational basis.

The communication channel for encoding is given by the four particle
GHZ-type maximally entangled state in the tilded basis
\begin{eqnarray}
\left \vert \xi_{e} \right \rangle =\frac{1}{2} \left ( \left
\vert {0} {0} {0} {0} \right \rangle + \left \vert {1} {1} {1} {1}
\right \rangle \right).
\end{eqnarray}
The sender and the receivers follow the information distribution
protocol. The sender performs the Bell-type measurement of the
input and port qubits and broadcasts the measurement result to the
receivers. Depending on the four different measurement outcomes
$\left \vert \Phi_{nm} \right \rangle$, the receivers perform the
LRUOs. Then information of the input qubit is encoded into the
three qubit state
\begin{eqnarray}
\left \vert \phi \right \rangle= \alpha_0 \left \vert
{0} {0} {0} \right \rangle+\alpha_1 \left
\vert {1} {1} {1} \right \rangle.
\end{eqnarray}

For decoding and error correction, we require all the encoded
qubits (which may have a phase error) to be at the same site of
the port qubit.  We exploit a maximally entangled state $\left
\vert \xi_d \right \rangle=\left( \left \vert {0} {0} \right
\rangle+\left \vert {1} {1} \right \rangle \right)/\sqrt{2}$ as
the quantum channel.  The RUOs are given by $U_{01}=\sigma_x$ and
$U_{10}=\sigma_z$.  We perform the extended Bell-type measurement
with the encoded qubits and the port qubit.  After an error
occurs, the encoded state is mapped to one of the four different
states orthogonal to each other depending on the error index
$\zeta (=0,1,2,3)$.  For each $\zeta$, we have one of four
different Bell measurement outcomes, therefore we have one
measurement outcome out of $16$ possible joint states.  These 16
joint states are equivalent to the 16 maximally entangle states
for the four qubit GHZ-type state.  We use the ``full'' Hilbert
space of four qubits for error correction.

If no error occurs, the extended Bell type measurement projects
onto one of the only four states $\left \vert \Phi^0_{nm} \right
\rangle = \left \vert \Phi_{nm} \right \rangle$ ($n,m$=0,1), the
same as in the standard teleportation scheme.  If a phase error
occurs in the $\eta$th qubit (out of the three qubits), the phase
error interchanges the state $\left \vert {0} \right \rangle
\leftrightarrow \left \vert {1} \right \rangle$ of the $n$th
qubit.  The extended Bell type measurement projects into the
$\left \vert \Phi^\zeta_{nm} \right \rangle$, which is different
from $\left \vert \Phi^\pm \right \rangle$ or $\left \vert
\Psi^\pm \right \rangle$. We perform the appropriate local
operation depending on $n$ and $m$ to the output qubit.  Then we
decode it back to the original input state $\left \vert \phi
\right \rangle$.

Now we investigate the quantum channel for error correction.  For
the GHZ type maximally entangled state of $N$ particles, there is
no direct entanglement between any two qubits.  If we trace out
any one of the qubits of the GHZ type state, the rest is in
complete mixture of the two orthogonal states consisting of $N-1$
qubits.  We have seen that quantum information is transmitted only
via an entangled channel in the previous sections. How can we
explain flow of quantum information in our error correction scheme
via entanglement?  In the information encoding process, quantum
information of the input state should not be transmitted into any
of the qubits.  However, the port qubit is maximally entangled
with all the three output qubits. From this fact, we may consider
that quantum information is transmitted via entanglement among the
three qubits and no information is implemented in the local state
of each qubit.

\begin{figure}[H]
 \begin{center} \epsfig{file=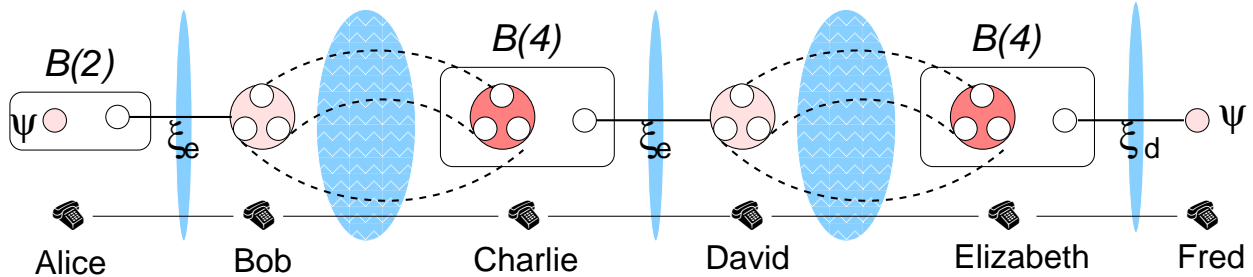,width=6.5truein}
\caption{The error correcting repeater using a combination of
the three qubit error-correction via entanglement.}
\label{fig:ecrepeater}
\end{center}
\end{figure}

The tele-error-correction scheme via entanglement channel is a
method for quantum communication that is secure against a single
qubit single type error attack of an eavesdropper.  The single
qubit single type attack of an eavesdropper appears in the encoded
state as an error. We can correct the error and retain information
of the input state. Thus the attack of the eavesdropper should not
gain any information of the input state.  We may consider an error
correction repeater using a combination of the
tele-error-correction schemes (Fig.~\ref{fig:ecrepeater}). Here we
present an example for an amplitude type attack, so we do not use
the tilded basis.  Consider Alice sending quantum information to
Fred. Bob, Charlie, David, Elizabeth are located between Alice and
Fred and pass through the quantum information.  Alice and Bob,
Charlie and David, Elizabeth and Fred are separated from each
other and connected via secure quantum channels. Bob and Charlie
are connected via an insecure channel and so are David and
Elizabeth, there may be a single error.  Their channels can be
non-perfect EPR pairs or even optical fibers with which one may
transmit a particle (photon).  Alice shares the quantum channel
for encoding $\left \vert \xi_e \right \rangle$ with Bob and so
does Charlie and David.  Elizabeth and Fred share a quantum
channel for decoding $\left \vert \xi_d \right \rangle$.  Alice
performs the Bell type measurement in the maximally entangled
state basis for two qubits denoted by ${\cal B} \left( 2 \right)$
and send 2-bit classical information to Bob.  Bob perform the
appropriate (L)RUO of his three qubits.  Bob sends information
from the encoded three qubits via the insecure channel.  Charlie
receives the three qubits from Bob.  An error might have happen to
one of the three qubits. Charlie performs the joint measurement on
the three qubits and the port qubit, which he shares with David in
the maximally entangled state basis of four qubits denoted by
${\cal B} \left( 4 \right)$. Charlie and David follow the protocol
of the information distribution scheme.  David sends his three
qubits via insecure channel to Elizabeth.  Finally Elizabeth
performs ${\cal B} \left (4 \right )$ together with her three
qubits and the port qubit of the quantum channel for decoding. The
information of the original state of Alice is now found at the
qubit hold by Fred.

We note that the distribution of quantum information for encoding
is similar to quantum secret sharing and splitting scheme
\cite{secretsharing,Cleve} if all the output state qubits are
spatially separated as pointed out in \cite{Cleve}.  In
\cite{Cleve}, Cleve et al stress that every quantum secret sharing
scheme is a quantum error-correcting code (in some sense), but
that error correction codes are not necessarily quantum secret
sharing codes.

\subsection{Correcting amplitude and phase errors}

Here we will show the output state basis for encoding in 5-qubit,
7-qubit, 9-qubit quantum error correction codes and their LRUOs.
Now the extended Bell type measurement involves 6 qubits and 8
qubits; it is almost impossible to distinguish all the different
outcomes, so it may not be practical, but it is interesting to
investigate entanglement of these quantum channels. As we will
show in the following, the quantum channel for encoding the
5-qubit code requires 3-ebit entanglement, the 7-qubit code
requires 2-ebit entanglement and the 9-qubit code requires 1-ebit
entanglement.  It is interesting that the most condensed error
correction code requires the most entanglement.

We first show the 7-qubit quantum error correction code via
entanglement because it contains higher symmetry.  The two output
state basis for encoding are
\begin{eqnarray}
\left \vert \phi_0
\right \rangle=\frac{1}{2 \sqrt{2}} \left \{
\left \vert 000 \right \rangle
\left( \left \vert0000 \right \rangle +
\left \vert 1111 \right \rangle \right)+
\left \vert 011 \right \rangle
\left( \left \vert0011 \right \rangle +
\left \vert 1100 \right \rangle \right)+ \right. \nonumber \\
\left.\left \vert 101 \right \rangle
\left( \left \vert0101 \right \rangle +
\left \vert 1010 \right \rangle \right)+
\left \vert 110 \right \rangle
\left( \left \vert 0110 \right \rangle +
\left \vert 1001 \right \rangle \right) \right\}, \\
\left \vert \phi_1
\right \rangle=\frac{1}{2 \sqrt{2}} \left\{
\left \vert 111 \right \rangle
\left( \left \vert0000 \right \rangle +
\left \vert 1111 \right \rangle \right)+
\left \vert 100 \right \rangle
\left( \left \vert0011 \right \rangle +
\left \vert 1100 \right \rangle \right)+  \right. \nonumber \\
\left. \left \vert 010 \right \rangle
\left( \left \vert0101 \right \rangle +
\left \vert 1010 \right \rangle \right)+
\left \vert 001 \right \rangle
\left( \left \vert 0110 \right \rangle +
\left \vert 1001 \right \rangle \right) \right \}.
\end{eqnarray}
The LRUOs are given by $U_{10}=\sigma_z \otimes \cdot \cdot \cdot
\otimes \sigma_z$ and $U_{01}=\sigma_x \otimes \cdot \cdot \cdot
\otimes \sigma_x$.  In this case, one of the even (or odd) order
of qubits will be the error corrected using the rest of the qubit.

The broadcasting channel for encoding $\left \vert \xi_e \right
\rangle=\frac{1}{\sqrt{2}} \left(\left \vert 0 \right \rangle
\left \vert \phi_0 \right \rangle +\left \vert 1 \right \rangle
\left \vert \phi_1 \right \rangle \right)$ can be written in the
following two ways:
\begin{eqnarray}
\left \vert \xi_e \right \rangle&=&\frac{1}{4}
\left \{
\left( \left \vert 0000 \right \rangle +
\left \vert 1111 \right \rangle
\right)
\left(\left \vert 0000 \right \rangle +
\left \vert 1111 \right \rangle
\right) \right. \nonumber \\
&+&
\left(\left \vert 0011 \right \rangle +
\left \vert 1100 \right \rangle
\right)
\left(\left \vert 0011 \right \rangle +
\left \vert 1100 \right \rangle
\right)  \nonumber \\
&+&
\left(\left \vert 0101 \right \rangle +
\left \vert 1010 \right \rangle
\right)
\left(\left \vert 0101 \right \rangle +
\left \vert 1010 \right \rangle
\right)  \nonumber \\
&+&
\left. \left(\left \vert 0110 \right \rangle +
\left \vert 1001 \right \rangle
\right)
\left(\left \vert 0110 \right \rangle +
\left \vert 1001 \right \rangle
\right) \right \} \\
&=& \left (\left \vert 00 \right \rangle + \left \vert 11\right \rangle \right)
\left (\left \vert 00 \right \rangle + \left \vert 11\right \rangle \right)
\left (\left \vert 00 \right \rangle + \left \vert 11\right \rangle \right)
\left (\left \vert 00 \right \rangle + \left \vert 11\right \rangle \right)
\nonumber \\
&+&
\left (\left \vert 00 \right \rangle - \left \vert 11\right \rangle \right)
\left (\left \vert 00 \right \rangle - \left \vert 11\right \rangle \right)
\left (\left \vert 00 \right \rangle - \left \vert 11\right \rangle \right)
\left (\left \vert 00 \right \rangle - \left \vert 11\right \rangle \right)
\nonumber \\
&+&
\left (\left \vert 01 \right \rangle + \left \vert 10\right \rangle \right)
\left (\left \vert 01 \right \rangle + \left \vert 10\right \rangle \right)
\left (\left \vert 01 \right \rangle + \left \vert 10\right \rangle \right)
\left (\left \vert 01 \right \rangle + \left \vert 10\right \rangle \right)
\nonumber \\
&+&
\left (\left \vert 01 \right \rangle - \left \vert 10\right \rangle \right)
\left (\left \vert 01 \right \rangle - \left \vert 10\right \rangle \right)
\left (\left \vert 01 \right \rangle - \left \vert 10\right \rangle \right)
\left (\left \vert 01 \right \rangle - \left \vert 10\right \rangle \right)
\end{eqnarray}
Both representations suggest that this state can be considered to be a
maximally entangled state of four levels.  (The first representation
is in two maximally entangled $4$-level particles and the second is in
four maximally entangled $4$-level particles.) Thus the state has
$\log_2 4=2$ e-bit entanglement

For the 5-qubit error correction code via entanglement, the output
state basis for encoding is given by (in the representation of Barenco
et al \cite{Barenco})
\begin{eqnarray}
\left \vert \phi_0 \right \rangle
&=&
\left \vert 000+111\right \rangle
\left \vert 00 \right \rangle
-
\left \vert 010+101 \right \rangle
\left \vert 11 \right \rangle \nonumber \\
&+&
\left \vert 001+110 \right \rangle
\left \vert 01 \right \rangle
+
\left \vert 011+100 \right \rangle
\left \vert 10 \right \rangle, \\
\left \vert \phi_1 \right \rangle
&=& -
\left \vert 000-111\right \rangle
\left \vert 11 \right \rangle
-
\left \vert 010-101 \right \rangle \left \vert 00 \right \rangle
\nonumber \\ &-& \left \vert 001-110 \right \rangle \left \vert 10
\right \rangle + \left \vert 011-100 \right \rangle \left \vert 01
\right \rangle.
\end{eqnarray}
In this case, the first qubit will be error corrected using the
rest of qubits.  The role of qubits are rather asymmetric in this
case.  The LRUOs are given by
\begin{eqnarray}
U_{10}=\sigma_x \otimes \sigma_x \otimes \sigma_x \otimes {\bf 1}
\otimes {\bf 1}
\end{eqnarray}
and
\begin{eqnarray}
U_{01}=-\sigma_z \otimes \sigma_z \otimes \sigma_x
\otimes \sigma_z \cdot \sigma_x \otimes \sigma_z \cdot \sigma_x.
\end{eqnarray}
The quantum channel for this case can be represented by the
maximally entangled state of an 8-level system, which contains 3
e-bits of entanglement.

For the 9-qubit error correction code, the encoding output state
basis are:
\begin{eqnarray}
\left \vert \phi_0 \right \rangle
&=&\left \vert 000+111 \right \rangle  \left \vert 000+111 \right \rangle
\left \vert 000+111 \right \rangle, \\
\left \vert \phi_1 \right \rangle &=&\left \vert 000-111 \right
\rangle  \left \vert 000-111 \right \rangle \left \vert 000-111
\right \rangle.
\end{eqnarray}
The LRUOs are given by $U_{10}=\sigma_x \otimes \cdots \otimes
\sigma_x$ and $U_{01}=\sigma_x \otimes \cdots \otimes \sigma_x$
(although these LRUOs are not unique).  In this case, any of the
qubits in the state can be error corrected using the rest of
qubits, so the role of each qubit is very symmetric.  The quantum
channel for encoding is represented by the maximally entangled
state of 2-level system, which suggests the amount of entanglement
is 1 e-bit.

\section{Summary} \label{sec:summary}

We have presented a generalization of quantum teleportation for
distributing quantum information of a $d$-level particle from a sender
to $M$ remote receivers via an initially shared multiparticle
entangled state.  The entangled state functions as a multiparty
quantum channel for distributing information.  This entangled state is
a maximally entangled state between the port particle of the sender
and the output particles of the receivers.  The existence of two LRUO
($U_{10}^{local}$ and $U_{01}^{local}$) for the output state basis is
essential for our information distribution to allow multiple
receivers at arbitrary locations.

We have presented optimal quantum information broadcasting of a
$d$-level particle, asymmetric telecloning of qubits, and
tele-error-correction as examples of the quantum information
distribution scheme. For the quantum information broadcasting, we
show the pure output state for $1 \rightarrow M$ optimal cloning
of $d$-level particles including ancillas.  This output state is a
physical implementation of the optimal cloning map presented by
Werner \cite{Werner}.  The investigation of the asymmetric
telecloning for qubits suggests that {\it quantum} information of
the input qubit is only transmitted by a {\it directly} entangled
channel.  The tele-error-correction scheme provides another
interpretation of quantum error correction from the viewpoint of
entanglement and allows an interesting observation of the amount
of entanglement required for the quantum channel for encoding.
This scheme can be used for secure communication.

\acknowledgements

The authors are grateful to J. Watson for his help during the
preparation of this manuscript. We are supported by the Special
Postdoctoral Researchers Program of RIKEN, the Leverhulme Trust,
the UK Engineering and Physical Sciences Research Council. Parts
of this work were completed during the Isaac Newton Institute
program ``Communication, Complexity and Physics of Information''
(1999) and has benefited from the ESF meeting of the QIT
programme.

\appendix
\section{}

We show a detailed derivation of Eq.~(\ref{eqn:decompxi}).  First
we show the case $i=0$ and then show the case $i \neq 0$. For the
case of $i=0$, the terms which give non-vanishing contribution of
the scalar product in Eq.(\ref{eqn:decompxi}) are the terms which
contains at least one $\left \{ 0 \right \}$ in the computational
basis representation. This requires $a_0=0$. Only the first
$n_{M-1}$ out of $n_M$ terms in the symmetrized states $\left
\vert \xi_k^M \right \rangle$ are $a_0=0$ and contribute in the
sum of Eq.(\ref{eqn:decompxi}).  For $0 \le k \le n_{M-1}-1$, a
symmetric state can be decomposed into the two parts
\begin{eqnarray}
\left \vert \xi_k^M \right \rangle &=&
\frac{1}{\sqrt{{\cal N} \left(\xi_k^M \right)}}
\left \vert
{\cal P}_M \left (
0,a_1,\cdot \cdot \cdot ,a_{M-1}
\right)
\right \rangle_{PA}
\nonumber \\
&=& \frac{1}{\sqrt{{\cal N} \left(\xi_k^M \right)}}
\left \vert 0 \right \rangle_{P} \otimes
\left \vert
{\cal P}_{M-1} \left (
a_1, \cdot \cdot \cdot, a_{M-1}
\right)
\right \rangle \nonumber \\
&+&
\frac{1}{\sqrt{{\cal N} \left(\xi_k^M \right)}}
\sum_{a_j \neq 0} \left \vert a_j \right \rangle \otimes
\left \vert
{\cal P}_{M-1} \left (
0,\cdot \cdot \cdot a_{j-1},a_{j+1}, \cdot \cdot \cdot,a_{M-1}
\right)
\right \rangle.
\end{eqnarray}
The scalar product is now given by
\begin{eqnarray}
\sum_{k=0}^{n_{M}-1} {}_P \left \langle 0
\right \vert \left. \xi_k^M \right \rangle_{PA}
\otimes
\left \vert \xi_k^M \right \rangle_{C}
&=&
\sum_{k=0}^{n_{M-1}-1} {}_P \left \langle 0
\right \vert \left. \xi_k^M \right \rangle_{PA}
\otimes
\left \vert \xi_k^M \right \rangle_{C}
\nonumber \\
&=&\sum_{k=0}^{n_{M-1}-1}
\sqrt{
\frac{{\cal N} \left(\xi_{k}^{M-1} \right)}
{{\cal N} \left(\xi_{k}^M \right)}}
\left \vert \xi_{{k}^{M-1}} \right \rangle
\otimes
\left \vert \xi_k^M \right \rangle_{C}
\end{eqnarray}
which is a special case of Eq.(\ref{eqn:decompxi}) with $k=g \left
(0,k^\prime \right)= k^\prime$.

For $i \neq 0$, the decomposition of the $M$ particle symmetrized state
in terms of the $M-1$ symmetrized state is
\begin{eqnarray}
\left \vert \xi_k^M \right \rangle &=&
\frac{1}{\sqrt{{\cal N} \left(\xi_k^M \right)}}
\left \vert
{\cal P}_M \left (
a_0,a_1,\cdot \cdot \cdot ,a_{M-1}
\right)
\right \rangle_{PA} \nonumber \\
&=& \frac{1}{\sqrt{{\cal N} \left(\xi_k^M \right)}}
\left \vert a_j=i \right \rangle_{P} \otimes
\left \vert
{\cal P}_{M-1} \left (a_0,\cdot \cdot \cdot,
a_{j-1}, a_{j+1},\cdot \cdot \cdot, a_{M-1}
\right)
\right \rangle_A \nonumber \\
&+&
\frac{1}{\sqrt{{\cal N} \left(\xi_k^M \right)}}
\sum_{a_j \neq i} \left \vert a_j \right \rangle \otimes
\left \vert
{\cal P}_{M-1} \left (
a_0,\cdot \cdot \cdot, a_{j-1},a_{j+1}, \cdot \cdot \cdot,a_{M-1}
\right)
\right \rangle_A.
\end{eqnarray}
Then the scalar product is
\begin{eqnarray}
\sum_{k=0}^{n_{M}-1} {}_P \left \langle i
\right \vert \left. \xi_k^M \right \rangle_{PA}
\otimes
\left \vert \xi_k^M \right \rangle_{C}
&=&
\sum_{k^{\prime}=0}^{n_{M-1}-1}
\sqrt{\frac{{\cal N} \left(\xi_{k^{\prime}}^{M-1} \right)}
{{\cal N} \left(\xi_{k}^M \right)}}
\left \vert \xi_{k^{\prime}}^{M-1} \right \rangle_{PA}
\otimes
\left \vert \xi_k^M \right \rangle_{C}
\nonumber \\
&=&\sum_{k^{\prime}=0}^{n_{M-1}-1}
\sqrt{\frac{{\cal N} \left(\xi_{k^{\prime}}^{M-1} \right)}
{{\cal N} \left(\xi_{g \left(i,k^{\prime} \right)}^M \right)}}
\left \vert \xi_{k^{\prime}}^{M-1} \right \rangle_{PA}
\otimes
\left \vert \xi_{g \left(i,k^{\prime} \right)}^M \right \rangle_{C}
\end{eqnarray}
since $k=g \left(i,k^{\prime} \right)$.

\end{document}